\documentclass[11pt,twoside,twocolumn,english,aps,manuscript]{revtex4}
\usepackage[T1]{fontenc}
\usepackage[latin1]{inputenc}
\usepackage{babel}
\usepackage{graphics}

\makeatletter

\providecommand{\LyX}{L\kern-.1667em\lower.25em\hbox{Y}\kern-.125emX\@}
\let\SF@@footnote\footnote
\def\footnote{\ifx\protect\@typeset@protect
    \expandafter\SF@@footnote
  \else
    \expandafter\SF@gobble@opt
  \fi
}
\expandafter\def\csname SF@gobble@opt \endcsname{\@ifnextchar[%]
  \SF@gobble@twobracket
  \@gobble
}
\edef\SF@gobble@opt{\noexpand\protect
  \expandafter\noexpand\csname SF@gobble@opt \endcsname}
\def\SF@gobble@twobracket[#1]#2{}

\usepackage{comment}

\makeatother
\begin{document}

\title{Disentangling the transfer and breakup contributions from the inclusive
\( ^{8} \)Li+\( ^{208} \)Pb reaction}

\author{A. M. Moro and R. Crespo }

\address{Departamento de F\' {\i}sica, Instituto Superior T\'{e}cnico, 1049--001 Lisboa, Portugal}

\author{H. García-Martínez, E. F. Aguilera and E. Martínez-Quiroz}

\address{Departamento del Acelerador, ININ, A.P. 18--1027, C. P. 11801, México,
D. F.}

\author{J. Gómez-Camacho}

\address{Departamento de FAMN, Universidad de Sevilla, Apdo. 1065, 41080 Sevilla, Spain}

\author{F. M. Nunes}

\address{Universidade Fernando Pessoa, Pra\c{c}a 9 de Abril, 4200, Porto, Portugal}

\begin{abstract}
An analysis of the \( ^{8} \)Li+\( ^{208} \)Pb reaction at energies
around the Coulomb barrier is presented. The study is focused on the
elastic and one-neutron removal channels. For the elastic scattering,
an optical model analysis of the experimental data is performed. The
observed \( ^{7} \)Li is interpreted as the superposition of the
one-neutron transfer reaction, \( ^{208} \)Pb(\( ^{8} \)Li,\( ^{7} \)Li)\( ^{209} \)Pb,
and the breakup reaction. The separate contribution of each one of
these processes has been calculated within the DWBA formalism. The
sum of both contributions explains adequately the experimental angular
distribution of \( ^{7} \)Li\emph{. }

%The effect of the transfer channel on the elastic angular distribution is also studied and discussed. 

\\
\\
Presentamos un an\'alisis de la reacción \( ^{8} \)Li+\( ^{208} \)Pb
a energías en torno a la barrera coulombiana. El estudio se centra
en \emph{}los canales de sección eficaz el\'astica y de separación
de un neutr\'on. Para la dispersión el\'astica, se ha efectuado un
an\'alisis de modelo óptico de los datos experimentales. El \( ^{7} \)Li
observado ha sido interpretado como una superposición de la reacción
de transferencia de un neutrón,\( ^{208} \)Pb(\( ^{8} \)Li,\( ^{7} \)Li)\( ^{209} \)Pb,
y la reacción de fragmentación. La distribución de cada uno de estos
procesos se ha calculado dentro del formalismo DWBA. La suma de ambas
contribuciones explica adecuadamente la distribución angular del \( ^{7} \)Li
medido. 

%Se estudia y discute tambi\'en el efecto del canal de transferencia en la distribuci\'on angular el\'astica. 

\emph{Descriptores:} reacciones con n\'ucleos poco ligados, an\'alisis
de modelo óptico y DWBA, dispersión el\'astica, reacciones de transferencia
y ruptura.


% 24.10.Eq: Coupled-channel and distorted-wave models

% 24.50.+g Direct reactions

% 25.60.-t Reactions induced by unstable nuclei, 25.60.Je Transfer reactions

% 25.70.Hi Transfer reactions
\end{abstract}
\maketitle

\section{Introduction}

During nearly two decades a considerable amount of experimental data
of nuclear reactions with radioactive beams has been accumulated.
These data have constituted our main source of knowledge of exotic
nuclei. Due to the experimental techniques used in the production
of these beams, most of these experiments correspond to medium and
high energies (tens or hundreds of MeV per nucleon). At low energies,
however, the scarce amount of data and the theoretical difficulties
turn the situation more uncertain. Many important questions concerning
the relevance of the different reaction mechanisms or the effect of
the weakly bound nature of exotic nuclei on the scattering observables
remain unanswered. For instance, a problem that has been subject of
exhaustive theoretical and experimental work \cite{Dia01,Hin02,Hus94,Das96,Yab97}
but does not seem to be fully clarified is the possible role of the
halo nucleon/s on the enhancement or suppression of the fusion cross
section in subbarrier reactions. Another question that requires further
analyses is the apparent absence of the threshold anomaly in weakly
bound systems \cite{Kim01}. Another striking result is the large
transfer/breakup cross section observed in the \( ^{6} \)He+ \( ^{209} \)Bi
reaction at energies below the barrier \cite{Agu00, Agu01}. 

In a recent experiment performed in the RNB facility at Notre-Dame,
the scattering of \( ^{8} \)Li on \( ^{208} \)Pb at bombarding energies
around the Coulomb barrier was measured \cite{Kol02}. This experiment
holds interesting repercussions within the context of the previoius
discussion, since it involves the weakly bound nucleus \( ^{8} \)Li.
As a distinctive feature, this reaction displays a prominent \( ^{7} \)Li
group. The large cross section associated with this group is interpreted
as an enhancement of the one-neutron removal channel due to the weak
binding of the valence neutron in \( ^{8} \)Li. This result resembles
the strong \( ^{4} \)He group observed in the \( ^{6} \)He+\( ^{209} \)Bi
reaction \cite{Agu00}. However, while in the \( ^{6} \)He case a
realistic description of the transfer and breakup mechanisms would
require a three-cluster model, in the case of the \( ^{8} \)Li projectile
a two-cluster model (\( ^{7} \)Li core plus a valence neutron) may
be adequate for this problem. Thus, it is expected that the analysis
of the \( ^{8} \)Li+\( ^{208} \)Pb reaction is significantly easier
and can permit the extraction of useful physical information about
the relevant reaction mechanisms.

%is believed to come from the single-neutron transfer reaction \( ^{208} \)Pb(\( ^{8} \)Li,\( ^{7} \)Li)\( ^{209} \)Pb or from the breaukp reaction, leading to the three-body %continuum \( n \)+\( ^{7} \)Li+\( ^{208} \)Pb. Since the experiment did not allow to measure the removed neutron, it was not possible to separate relative importance %contributions. 

In this work, we present an analysis of the referred experiment, with
special emphasis on the elastic and single-neutron removal channels.
We first study the elastic angular distributions, in terms of optical
potentials. From this analysis we can conclude on the relevance of
non-elastic channels. The obtained optical potentials will also be
used in the analysis of the transfer and breakup channels. For these,
DWBA calculations will be performed. 

%Finally, the interplay between the elastic and transfer channels is studied and discussed.

\section{Analysis of elastic scattering data}

The study of the elastic scattering by means of optical potentials
(OP) provides information on the nature of the interaction between
the colliding nuclei and on the importance of non-elastic channels. 

In previous works \cite{Smi91, Bec93}, the similitude between the
scattering of \( ^{8} \)Li and \( ^{7} \)Li has been made apparent.
In these works, it is shown that the same potentials that fit the
\( ^{7} \)Li scattering on \( ^{12} \)C and \( ^{58} \)Ni give
also a good agreement for the \( ^{8} \)Li scattering on the same
targets and at similar energies.

\begin{comment}

For this purpose, we performed and optical model analysis of the elastic
data, using the OP for \( ^{7} \)Li+\( ^{208} \)Pb at \( E \)=33~MeV
obtained in a previous analysis \cite{Ism95}. We found that the original
parameters clearly overestimates the data at angles around the rainbow.
Then, we performed a best-fit search, using the \( ^{7} \)Li+\( ^{208} \)Pb
OP as initial parameters. We found that, varying the imaginary depth
and diffuseness, a very good agreement with the experimental angular
distributions was obtained, as can be observed in Fig.\ \ref{Fig:elastic_all}.
Nevertheless, we found that the resulting OP are very absorptive.
This is a clear indication of the importance of reaction channels
in the \( ^{8} \)Li reaction, such as the one-neutron removal, that
will be treated below.

\end{comment}
\begin{figure}
{\centering \resizebox*{1\columnwidth}{!}{\includegraphics{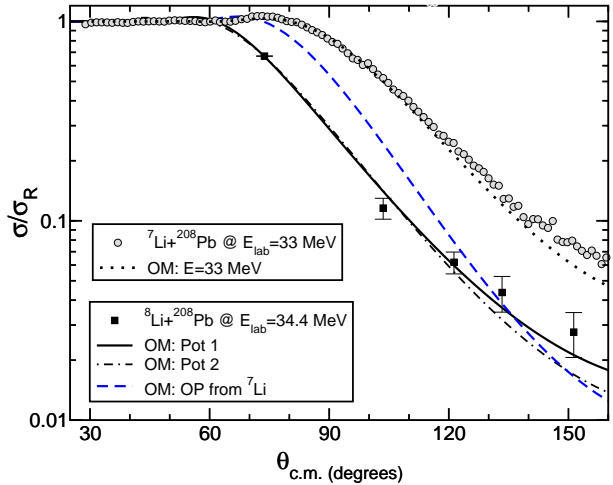}} \par}

\caption{\label{Fig:li8-li7}Comparative study for the elastic scattering
of \protect\( ^{8}\mathrm{Li}\protect \) and \protect\( ^{7}\protect \)Li
on \protect\( ^{208}\protect \)Pb. See text and labels for details.}
\end{figure}

We examine this result in the present reaction. For this purpose,
we compare in Fig.~\ref{Fig:li8-li7} the experimental data for \( ^{8} \)Li+\( ^{208} \)Pb
at 34.4~MeV with the data for \( ^{7} \)Li+\( ^{208} \)Pb at \( E_{lab} \)=33~MeV,
reported in \cite{Kee94}. The latter can be accurately described
using the OP listed in Table \ref{Tab:li8-li7} and represented by
the dotted line in Fig.~\ref{Fig:li8-li7} . Using these parameters
we performed an OM calculation for \( ^{8} \)Li+\( ^{208} \)Pb at
\( E_{lab} \)=34.4~MeV. This calculation, shown by the dashed line
in Fig.~\ref{Fig:li8-li7}, clearly fails to reproduce the data, particularly
at angles around the rainbow. The quality of the fit can be improved
by using a more absorptive potential. In Fig.~\ref{Fig:li8-li7} we
show the result obtained with two different optical potentials: the
first one, labeled as {}``Pot 1'' in Table \ref{Fig:li8-li7} and
plotted with the solid line, was obtained by fitting the depths and
the imaginary diffuseness to the data. This procedure leads to a very
deep imaginary potential. Due to the strong absorption, the calculations
are not very sensitive to the real potential. We found that keeping
the same real part of the interaction as that for \( ^{7} \)Li+\( ^{208} \)Pb,
a good fit to the experimental data could also be achieved using a
diffuse imaginary potential ({}``Pot 2'' in Table \ref{Fig:li8-li7}).
The corresponding elastic angular distribution is given by the dotted-dashed
line in Fig.~\ref{Fig:li8-li7}. These calculations suggest a more
absorptive nature in the case of \( ^{8} \)Li, in apparent contradiction
with previous results \cite{Smi91, Bec93}. This discrepancy might
be explained by noting that at energies around the barrier the OP
depends significantly on the energy and also that the \( ^{8} \)Li
and \( ^{7} \)Li barriers differ by about 4~MeV \cite{Kol02}.

It is worthwhile to note that the reported elastic data contains a
contribution from the \( 1^{+} \) excited state of the \( ^{8} \)Li
nucleus (\( \epsilon _{x} \)=0.98~MeV), which could not be separated
in the experiment. This inelastic contribution can be extracted in
a model dependent way by performing a coupled channels (CC) calculation
in which the ground and excited states are included explicitly, and
coupled to all orders. For the coupling matrix elements we took the
reported values for the electric transition probability, \( B(E2;2^{+}\rightarrow 1^{+}) \),
and the deformation parameter, \( \beta  \) \cite{Smi91}. Further
details of this calculation are given in \cite{Mor03a}. The main conclusion
is that there is an important contribution from inelastic events at
large angles.

We extended the optical model analysis to lower energies. The extracted
optical potential parameters, given in \cite{Mor03a}, indicate that
very absorptive potentials are also required at energies down to \( E_{cm} \)=24.4~MeV.

\begin{comment}

\begin{figure}
{\centering \resizebox*{1\columnwidth}{!}{\includegraphics{elastic_all.eps}} \par}

\caption{\label{Fig:elastic_all}Elastic cross section angular distribution,
as ratio to Rutherford cross section, for the scattering of \protect\( ^{8}\protect \)Li
on \protect\( ^{208}\protect \)Pb. }
\end{figure}

\end{comment}
\begin{table}

\caption{\label{Tab:li8-li7}Optical potential parameters used in the analysis
of \protect\( ^{7}\protect \)Li+\protect\( ^{208}\protect \)Pb elastic
scattering at \protect\( E_{lab}\protect \)=33~MeV and \protect\( ^{8}\protect \)Li+\protect\( ^{208}\protect \)Pb
at \protect\( E_{lab}\protect \)=34.4~MeV. In all cases, \protect\( r_{0}=r_{w}\protect \)=1.3~fm,
\protect\( a_{0}\protect \)=0.65~fm and \protect\( r_{C}=1.25\protect \)~fm.
The last column gives the chi square per data point.}

{\centering \begin{tabular}{cccccc}
\hline 
&
\( V_{0} \)&
\( W \)&
\( a_{w} \)&
\( \sigma _{R} \)&
\( \chi ^{2} \)\\
&
(MeV)&
(MeV)&
(fm)&
(mb)&
\\
\hline 
\( ^{7} \)Li+\( ^{208} \)Pb&
15.4&
13.2&
0.65&
&
-\\
\( ^{8} \)Li+\( ^{208} \)Pb (Pot 1)&
37.1&
154&
0.57&
1231&
1.11\\
\( ^{8} \)Li+\( ^{208} \)Pb (Pot 2)&
15.4&
58.4&
0.70&
1275&
1.51\\
\hline
\end{tabular}\par}
\end{table}

\section{The transfer/breakup reaction}

As observed in previous reactions involving \( ^{8} \)Li as a projectile,
the present experiment shows a large cross section for the \( ^{7} \)Li
yield, coming from the one-neutron removal channel. This phenomenon
is a consequence of (i) the weak binding of the valence neutron (\( \epsilon _{0} \)=2.03~MeV),
which results in a positive \( Q \) value, and (ii) the \( 2^{+} \)
spin of the \( ^{8} \)Li ground state, which allows multiple angular
momentum transfers. According to \( Q \) value arguments, both the
one-neutron and breakup channels are energetically favoured so, in
principle, both processes can contribute substantially to the measured
\( ^{7} \)Li yield. 

%[Since the valence neutron is only bound by \( \epsilon _{0}\approx  \)2 MeV, the \( ^{8} \)Li projectile can be easily broken up due to the tidal forces induced by %the target, and so the breakup channel is also expectoed to contributed significantly to the measured \( ^{7} \)Li yield.

One of the objectives of this work is to evaluate the relative contributions
of both mechanisms, and to verify that their sum explains the total
measured yield.

\subsection{Single-neutron transfer}

\begin{figure}
{\centering \resizebox*{1\columnwidth}{!}{\includegraphics{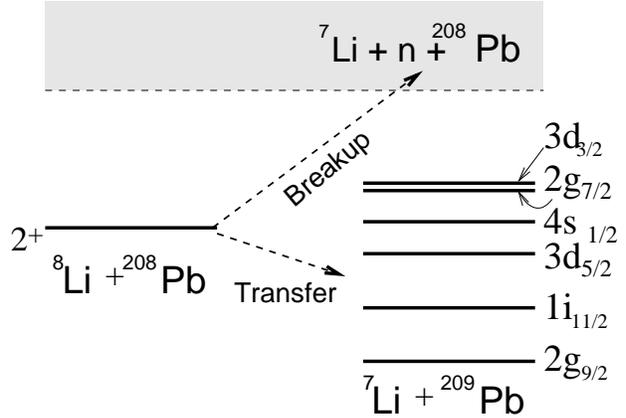}} \par}

\caption{\label{Fig:transbu-coup} Coupling scheme used in the DWBA calculations
for one-neutron transfer and breakup. }
\end{figure}

Within the post form of the Distorted Wave Born Approximation (DWBA),
the one-neutron transfer amplitude for our reaction reads \begin{equation}
\label{Eq:DWBA}
T=\langle \chi _{[^{7}\mathrm{Li},^{209}\mathrm{Pb}]}\phi _{[n,^{208}\mathrm{Pb}]}|V_{trans}|\chi _{[^{8}\mathrm{Li},^{208}\mathrm{Pb}]}\phi _{[n,^{7}\mathrm{Li}]}\rangle 
\end{equation}
 with the transfer operator \( V_{trans}=V_{[n,^{7}\mathrm{Li}]}+U_{[^{7}\mathrm{Li},^{208}\mathrm{Pb}]}-U_{[^{7}\mathrm{Li},^{209}\mathrm{Pb}]} \).
For the interaction \( V_{[n,^{7}\mathrm{Li}]} \), that binds the
neutron to the \( ^{7} \)Li core, we took the \( ^{8} \)B binding
potential used in \cite{Esb96}, with the depth modified to reproduce
the binding energy. In Eq.\ (\ref{Eq:DWBA}), \( \chi _{[^{8}\mathrm{Li},^{208}\mathrm{Pb}]} \)
and \( \chi _{[^{7}\mathrm{Li},^{209}\mathrm{Pb}]} \) are the distorted
waves describing the elastic scattering in the entrance and exit channels,
respectively. The former was generated with the potential {}``Pot
1'' of Table \ref{Tab:li8-li7}, and the latter with the \( ^{7}\mathrm{Li}+^{209}\mathrm{Pb} \)
optical potential, also given in this table. The functions \( \phi _{[n,^{208}\mathrm{Pb}]} \)
and \( \phi _{[n,^{7}\mathrm{Li}]} \) represent the overlap integrals
(\( ^{209} \)Pb|\( ^{208} \)Pb) and (\( ^{8} \)Li|\( ^{7} \)Li)
between internal states . The transfer was calculated to all bound
states in \( ^{209} \)Pb (see Fig.~\ref{Fig:transbu-coup}). These
states were described using pure single-particle configurations, with
the quantum numbers and spectroscopic factors taken from the literature.
In the case of \( ^{8} \)Li, we assume a core plus valence neutron
structure with the components \begin{eqnarray}
\rm {(a)} &  & \left[ |^{7}\mathrm{Li}(3/2^{-})\rangle \otimes |p_{3/2}\rangle \right] _{2^{+}},\nonumber \\
\rm {(b)} &  & \left[ |^{7}\mathrm{Li}(3/2^{-})\rangle \otimes |p_{1/2}\rangle \right] _{2^{+}},\nonumber \\
\rm {(c)} &  & \left[ |^{7}\mathrm{Li}(1/2^{-})\rangle \otimes |p_{3/2}\rangle \right] _{2^{+.}}\label{Eq:8Liwf} 
\end{eqnarray}

Therefore, the relevant spectroscopic amplitudes for the ground state
of \( ^{8} \)Li are \( A_{3/2}=CFP(2^{+}|3/2^{-},1p_{3/2}) \), \( A_{1/2}=CFP(2^{+}|3/2^{-},1p_{1/2}) \),
and \( B_{3/2}=CFP(2^{+}|1/2^{-},1p_{3/2}) \), where we have used
the definition

\begin{equation}
CFP(J|J_{7},nlj)=\frac{\langle ^{8}\mathrm{Li}(J)||a_{nlj}^{+}||^{7}\mathrm{Li}(J_{7})\rangle }{\sqrt{2J+1}},
\end{equation}
with the \( 3j \) phase convention of \cite{Tal93}. 

\begin{comment}

\begin{eqnarray}
|^{8}\mathrm{Li}\rangle _{gs} & = & A_{3/2}\left[ |^{7}\mathrm{Li}(3/2^{-})\rangle \otimes |p_{3/2}\rangle \right] _{2^{+}}\nonumber \\
 & + & A_{1/2}\left[ |^{7}\mathrm{Li}(3/2^{-})\rangle \otimes |p_{1/2}\rangle \right] _{2^{+}}\nonumber \\
 & + & B_{3/2}\left[ |^{7}\mathrm{Li}(1/2^{-})\rangle \otimes |p_{3/2}\rangle \right] _{2^{+}}\, .\, \, \, \, \, \label{Eq:8Liwf} 
\end{eqnarray}

The coefficients \( A_{3/2} \), \( A_{1/2} \) and \( B_{3/2} \)
are the spectroscopic amplitudes for the \( (^{8} \)Li|\( ^{7} \)Li)
overlap.

\end{comment}

We adopted the values: \( A_{3/2}=\sqrt{8/9} \), \( A_{1/2}=\sqrt{2/9} \)
and \( B_{3/2}=-\sqrt{2/9} \), which are derived in Ref. \cite{Mor03a}
making use of Wigner supermultiplet scheme. Note that the component
(c) in (\ref{Eq:8Liwf}) corresponds to a \( ^{8} \)Li configuration
with the \( ^{7} \)Li core in the excited state \( 1/2^{-} \) at
\( \epsilon _{x} \)=0.48~MeV. Since the energy resolution of the
present experiment did not permit the separation of both states, this
component has to be included in the calculation of the \( ^{7} \)Li
cross section. The calculated cross sections are shown in Fig.~\ref{Fig:li7ang}.
The thick dashed line represents the so called elastic transfer, where
the \( ^{7} \)Li is emitted in the ground state. The thin dashed
line is the inelastic transfer, coming from the component (c) in Eq.~(\ref{Eq:8Liwf}).
We see that the elastic transfer accounts for most of the observed
\( ^{7} \)Li, while the inelastic transfer gives a small (but non
negligible) contribution.

\subsection{The breakup reaction}

The breakup process can be visualized as a transfer of the valence
neutron to the target continuum (see Fig.\~{ }\ref{Fig:transbu-coup}):
\( ^{208} \)Pb(\( ^{8} \)Li,\( ^{7} \)Li)\( n \)\( ^{208} \)Pb.
Within this approach, the breakup cross section can be calculated
using the expression (\ref{Eq:DWBA}), just by replacing the \( ^{209} \)Pb
bound state wave functions by \( n \)-\( ^{208} \)Pb scattering
functions. Since the continuum spectrum contains an infinite set of
states in both energy and relative angular momentum, a procedure of
discretization was performed. For this purpose, for each partial wave
the continuum was divided into \( N=3 \) energy bins of equal width
and up to \( \epsilon =6 \)~MeV for the maximum relative energy
between \( n \) and \( ^{208} \)Pb. Only the partial waves \( 0\leq \ell \leq 6 \)
were retained in the calculation. Test calculations showed that the
contribution from higher partial waves gives a small contribution
to the breakup. As in the case of the transfer, the final \( ^{7} \)Li
can be produced in the ground or excited state, giving rise to \emph{elastic}
and \emph{inelastic} breakup contributions, respectively.
\begin{figure}
{\centering \resizebox*{1\columnwidth}{!}{\includegraphics{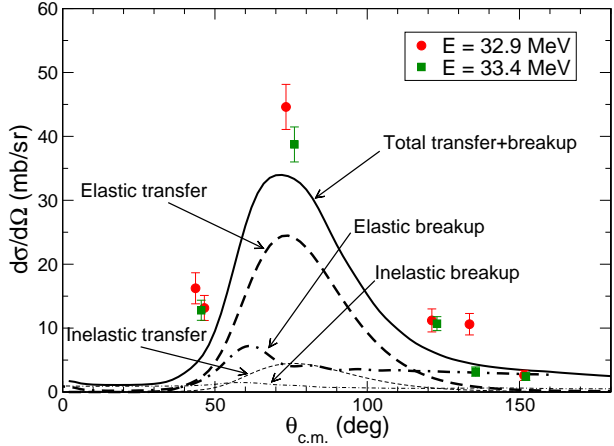}} \par}

\caption{\label{Fig:li7ang}Angular distribution for the \protect\( ^{7}\protect \)Li
yield produced in the \protect\( ^{8}\protect \)Li+ \protect\( ^{208}\protect \)Pb
reaction at \protect\( E_{lab}\protect \)=34.4 MeV. The dots and
squares correspond to the experimental data of Ref.~\cite{Kol02}.
The lines correspond to our DWBA calculations for the transfer and
breakup contributions, as indicated by the labels. }
\end{figure}

The angular distributions calculated for the elastic and inelastic
breakup cross sections are represented in Fig.~\ref{Fig:li7ang} by
the thick and thin dotted-dashed lines, respectively. Also shown is
the sum of the calculated transfer and breakup contributions, given
by the thick solid line. As it can be seen, it gives a fairly good
agreement with the experimental data, reproducing the shape of the
angular distribution and its normalization. According to our calculations,
the transfer channel accounts for about 80\% of the observed \( ^{7} \)Li
yield. However, it is interesting to note that for scattering angles
above 120 degrees, there is an inversion of the relative importance
of both mechanisms. In this region, the \( ^{7} \)Li particles are
mainly produced through the elastic breakup of the projectile.

\section{Summary and conclusions}

In summary, we have analyzed the recently measured reaction \( ^{8} \)Li+\( ^{208} \)Pb
at energies slightly above the Coulomb barrier. Experimentally, this
reaction shows, as a remarkable feature, a strong yield of \( ^{7} \)Li
particles, which is attributed to the weak binding of the valence
neutron in the \( ^{8} \)Li nucleus. An optical model analysis of
the elastic differential cross sections reflects a very absorptive
nature for the process, probably due to the loss of flux in the elastic
channel leading to the production of \( ^{7} \)Li and \( ^{4} \)He.
Our DWBA calculations indicate that the transfer cross section is
the main mechanism responsible for the one-neutron removal channel,
although there is also an important contribution coming from the breakup
channel, which is actually the dominant mechanism at backward angles.
The sum of both contributions accounts quite well for the measured
\( ^{7} \)Li angular distribution in both shape and absolute magnitude.
Other effects, such as the the importance of the \( 1^{+} \) excited
state on the transfer cross section and the effect of the transfer
channels on the elastic cross sections are now under study and will
be published elsewhere \cite{Mor03a}.

\begin{acknowledgments}
We are grateful to A. Mukhamedzhanov and N. Timofeyuk for useful discussions
concerning the \( ^{8} \)Li structure. This work has been partially
supported by the Ministerio de Ciencia y Tecnología under project
FPA2002-04181-C04-04 (Spain), Funda\c c\~ao para a Ci\^encia e a Tecnologia
(F.C.T.) under grant POCTIC/36282/99 (Portugal) and CONACYT (México).
A.M.M acknowledges a postdoctoral grant from F.C.T.
\end{acknowledgments}
\bibliographystyle{apsrev}
\bibliography{refer}

\end{document}